# *Ab-initio* investigation of the interfacial structural, electronic, and magnetic properties for $Co_2MnAl/X$ (X = MgO and GaAs) heterostructures

Amar Kumar[1], Mitali[1], Sujeet Chaudhary[1]*, and Sharat Chandra[2]*

[1]Thin Film Laboratory, Department of Physics, Indian Institute of Technology Delhi, New Delhi 110016, India

[2]Material Science Group, Indira Gandhi Centre for Atomic Research (a *CI* of Homi Bhabha National Institute), Kalpakkam, Tamil Nadu-603102, India

*Corresponding Authors: sujeetc@physics.iitd.ac.in and sharat@igcar.gov.in

## Abstract

The structural, electronic, and magnetic properties of (100)-oriented $Co_2MnAl/MgO$ and $Co_2MnAl/GaAs$ heterostructures are investigated using plane-wave pseudopotential density functional theory. For the $Co_2MnAl/MgO$, CoCo-MgMg, CoCo-OO, MnAl-MgMg, and MnAl-OO interfaces in top-to-top configurations are studied, while for $Co_2MnAl/GaAs$, both top-to-top (Co-Ga, Co-As, Mn-Ga, Mn-As, Al-Ga, Al-As) and bridge-site (CoCo-Ga, CoCo-As, MnAl-Ga, MnAl-As) interfaces are considered. The interfacial geometries featuring Co- or CoCo-atomic terminations for the $Co_2MnAl$ slab exhibit larger adhesion energies compared to those terminated with Mn-, Al-, or MnAl-atomic terminations. This indicates their greater interfacial stability. In contrast, MnAl-, Mn-, or Al-terminated interfaces preserve near half-metallicity, whereas Co- and CoCo-terminated geometries display a strongly metallic character. All studied interfaces show enhanced magnetic moments relative to their bulk counterparts, primarily arising from interfacial atoms and their nearest neighbours. These findings offer valuable insights for optimizing Co2MnAl-based heterostructures in spintronic applications.

# *Ab-initio* investigation of the interfacial structural, electronic, and magnetic properties for $Co_2MnAl/X$ (X = MgO and GaAs) heterostructures

Amar Kumar[1], Mitali[1], Sujeet Chaudhary[1]*, and Sharat Chandra[2]*

[1]Thin Film Laboratory, Department of Physics, Indian Institute of Technology Delhi, New Delhi 110016, India

[2]Material Science Group, Indira Gandhi Centre for Atomic Research (a *CI* of Homi Bhabha National Institute), Kalpakkam, Tamil Nadu-603102, India

*Corresponding Authors: sujeetc@physics.iitd.ac.in and sharat@igcar.gov.in

## I. Introduction

The next-generation spintronic devices are based on the multi-layered structures made of various magnetic and non-magnetic materials. Among these multilayered structures, magnetic tunnel junctions (MTJs) and spin valves (SVs) are particularly significant, serving as the foundational components in many advanced spintronic devices, such as magneto-resistive random-access memories (MRAMs), spintronic sensors, read heads of hard disk drives, spin transfer torque oscillators, logic devices, *etc*. Notably, the key component for both MTJs and SVs is the ferromagnetic/non-magnetic (FM/NM) heterostructure, and the performance of spintronic devices is strongly dependent on the selection of the composing layers, as noted from the literature [1].

For the FM/NM heterostructure for MTJs or SVs, FM materials with high spin polarization and high Curie temperatures (in their bulk form) are obviously preferred, as they can easily facilitate the generation, manipulation, and transport of spin currents. Conventional FM choices include the alloys of FM metals (Fe, Co, Ni), alloys of Co, Fe, and B, perovskite materials, and Heusler alloys (HAs). Although widely explored, these material classes possess certain intrinsic limitations for some key physical properties that hinder their applicability across a broad range of spintronic devices. However, with the continuous advancement of experimental techniques, HAs have emerged as a more promising and versatile class of materials for wide spintronic applications. HAs accommodated a large family of ~1500 multifunctional compounds, which offers a broad range of tunable and device-specific spintronic properties across many of their members. These properties include low magnetization, reduced magnetic damping, high magnetic anisotropy, and significant magnetoresistance, while still meeting the fundamental requirements for spintronic applications - namely, ferromagnetic ordering and high Curie temperatures [2,3]. For instance, MTJs incorporating HAs as FM electrodes have demonstrated higher tunnelling magnetoresistance (TMR) ratios, which is an indispensable parameter for measuring the transport performance of MTJs; compared to those utilizing other FM materials, such as - FM metal alloys, CoFeB, or perovskite materials. The current highest experimentally observed TMR values for MTJs with using these above mentioned FM materials are as:

1995% (measured at 4 K) for $Co_2MnSi/MgO/Co_2MnSi$ MTJs based on HAs [4], 1144% (measured at 5 K) for CoFeB/MgO/CoFeB MTJs based on CoFeB alloy [5], 1143% (measured at 10 K) for CoFe/MgO/CoFe(001) MTJs based on FM metals alloys [6], and 180% (measured at 180K) for $La_{0.7}Sr_{0.3}MnO_3/BaTiO_3$-$La_{0.5}Ca_{0.5}MnO_3/La_{0.7}Sr_{0.3}MnO_3$ MTJs based on perovskite materials [7]. These observations indicate that FM/NM heterostructures incorporating HAs as the FM layer may exhibit superior spintronic performance compared to those employing other FM materials. Regarding the choice of NM material, the selection of a particular NM layer in a HM/NM heterostructures depends on the several factors beside the electronic properties, such as - structural compatibility with the adjacent FM layers (*i.e.*, efficiency of making high symmetrical heterostructures with closely matched lattice structures), interfacial stability with the adjacent FM materials, and their physical properties, such as - crystal structure, spin transport, electrical properties, *etc*. The selected NM is often the insulating material for MTJs or the semiconducting material for SVs [8,9]. Thereby, the FM/NM heterostructures comprising various FM materials belonging to the HAs family and different NM materials hold significant potential for advancing spintronic applications. To be more precise, beyond the careful selection of FM and NM layers, the experimental efficiency of FM/NM heterostructures is found to be highly sensitive to interfacial morphologies of the constituent layers and can significantly influence the physical properties of heterostructures. Therefore, a detailed investigation of the interfacial geometry-dependent properties of FM/NM heterostructures is essential prior to device design to ensure optimal device efficiency. In this context, the present study focuses on the various physical properties related to the interfacial geometries of FM/NM heterostructures, with the FM material belonging to HAs. The particular selection of FM and NM materials is made based on the following considerations:

Among the large family of HAs, Co-based Heusler compound $Co_2MnAl$ has been chosen as FM in this study, due to its well-established and versatile spintronic properties. These include a nearly half-metallic nature (with ~75% theoretical spin polarization [10], ~67% experimental spin polarization [11]), a moderate magnetization ~ 4.01 $\mu_B$/*f.u.* *[10]*), high curie temperature ~ 720 K [10], structural imperfections resilience nature for the electronic band structure and magnetization [10,12,13], very large anomalous Hall conductivity of ~ 1421.6 S/cm, and very large spin Hall conductivity of ~ 694.8 $\hbar$/e S/cm (both AHC and SHC are the largest among the Co-based full HAs) [14], a record anomalous hall angle (representing transverse charge accumulation efficiency) of ~ 0.21 [15], a large spin hall angle (representing spin to charge conversion efficiency) of ~ 0.08 [14].

However, despite such advantageous spintronic properties of $Co_2MnAl$, there is a notable lack of detailed studies on the interfacial properties of $Co_2MnAl$/NM heterostructures in the literature. This includes fundamental spintronic characteristics such as interfacial magnetization, electronic band structures, spin polarization, and magnetic anisotropy. The only relevant study, conducted by Sakuma *et al*. [16], primarily explores the electronic nature and interfacial magnetization for one interfacial geometry in Co$_2$MnAl/MgO heterostructures. Thus, literature clearly demands the study of the physical properties of Co$_2$MnAl/NM heterostructures. For the NM layer, MgO and GaAs are proven to the excellent choices when Co-based full HAs are selected as the FM component in HA/NM heterostructures for MTJs or SVs. Particularly, MgO is the most widely used tunnel barrier in MTJs [8], whereas GaAs is commonly employed as NM material in SV structures [9] [10.1103/RevModPhys.76.323]. Also, the MgO and GaAs have a very small lattice mismatch with Co$_2$MnAl. Considering all these factors, in the present study, the various physical properties for various interfacial morphologies within Co$_2$MnAl/X /X heterostructures (X=MgO and GaAs) are studied, using first-principles-based density functional theory. In this context, first-principles calculations are highly advantageous and play a vital role by providing atomistic insights and guiding material design, without the need for expensive experimental trials. All heterostructures in the present study are modeled considering the ideal bulk structure for constituting layers, *i.e.*, L2$_1$-ordered structure for Co$_2$MnAl, *rock-salt* structure for MgO, and *zinc blende* structure for GaAs, and considering the ideal epitaxies for the modelled interfacial morphologies. For heterostructures modelling, the [001] orientation for all slabs is selected in the out-of-plane direction.

Noticeably, a more rigorous study, combining the different interfacial geometries with structural imperfections, could be closer to the real experiments; however, due to the computational simplicity, we are limited to the interfaces combining epitaxial geometries. Furthermore, to ensure the later experimental utilization of these results, the thicknesses of Co$_2$MnAl, MgO, and GaAs layers are chosen within the experimentally appropriate regime for modeling interfaces. The justifications behind the thickness choices are also discussed later in the result and discussion section (see the interface modelling Subsection). Since the interface morphologies have been modeled using symmetric slabs of FM and NM with the equivalent terminations at both ends, due to thickness choice within practical range; the results of this study can be utilized in two different ways – (i) to study Co$_2$MnAl thin film(s)' physical properties deposited on the MgO or GaAs substrate or buffer layer, and (ii) to get the evidence for iso-terminated MTJs/SVs with the Co$_2$MnAl/MgO/Co$_2$MnAl and Co$_2$MnAl/GaAs/Co$_2$MnAl configurations, as these heterostructures contain two equivalent interfaces (see the interface modelling Subsection). Further details about the interface modeling are provided in the interface modelling Subsection under Section-III. Additionally, Section-III enclosed the results for the structural, electronic, and magnetic properties of Co$_2$MnAl/NM heterostructures. Finally, all results are summarized in the concluding section- Section IV.

## II. Computational details

To study the physical properties of Co$_2$MnAl/X (X = MgO and GaAs) heterostructures, first-principles based density function theory (DFT) calculations are performed using the plane-wave-pseudopotential (PWP) method as implemented in QUANTUM ESPRESSO package [17,18]. The generalized gradient approximation (GGA), in the parameterization of Perdew–Burke–Ernzerhof (PBE) functional, is used to deal with the electronic exchange-correlation (XC) interactions. Although GGA+U is generally preferred for studying the physical properties of transition metal systems, the GGA functional has been adopted in this study, as it is reported to yield more consistent results with experimental observations for $Co_2MnAl$, compared to GGA+U in many previous studies [12,19–21]. For the Pseudopotentials, scalar relativistic pseudopotentials from the publicly available repository of the *QUANTUM ESPRESSO* -*"PSlibrary"* with the valence-electrons configurations of Co ($3s^23p^64s^23d^7$), Mn ($3s^23p^64s^23d^5$), Al ($3s^23p^1$), Mg ($2s^22p^63s^2$), O($2s^22p^4$), Ga($3d^{10}4s^24p^1$), and As ($4s^24p^3$) are used to handle with electron-ion interaction [22]. Here, the spin-orbit coupling (SOC) is not included in the calculations, as the SOC has a negligible impact on the electronic band structure and magnetization of bulk-$Co_2MnAl$ [23–25]. All heterostructures are geometrically optimized by allowing the relaxation of the atomic positions of five atomic layers of $Co_2MnAl$ and two layers of MgO or GaAs, along the *z* direction, and minimizing the Hellmann-Feynman forces below the $10^{-3}$ Ry/Bohr. A cut-off energy of 150 Ry for the plane wave expansion is used to describe the atomic basis set. To achieve electronic self-consistency, the convergence criteria for the total energy is set to $10^{-6}$ Ry. For the structural relaxation, a *k*-point mesh equivalent to 11×11×11 for the $L2_1$-ordered structure [(5.69)3 Å3 unit-cell volume]; while for the density of states calculations, a denser k-point mesh for the interfaces, equivalent to 15×15×15 for the $L2_1$-ordered structure, has been employed.

### III. Results and Discussion

**Interface Modelling:**

The epitaxial $Co_2MnAl$/X (X = GaAs/MgO) heterostructures are simulated using a (1×1) supercell, and contain two identical interfaces with identical atomic configurations. For interface modelling, the (100) orientation is considered for both layers– $Co_2MnAl$ and X (GaAs/MgO), *i.e.,* the $Co_2MnAl$(100)/MgO(100) and $Co_2MnAl$(100)/GaAs(100) heterostructures are considered. All heterostructures are modeled using the ideal bulk structures for the constituent layers: the $L2_1$-ordered structure for $Co_2MnAl$ with an optimized lattice parameter (OLP) of 5.69 Å, the *rock-salt* structure for MgO with an OLP of 4.25 Å, and the *zinc-blende* structure for GaAs with an OLP of 5.74 Å. Given the layered structure of $Co_2MnAl$, MgO, and GaAs along (100) orientation, numerous possible interfacial geometries can exist with the formation of a heterostructure. Accordingly, for systematic categorization, the interfacial structures are defined by referring to the atomic bonds (or connected atoms) at the interfaces, between the one atomic layer each from the HA and X slabs, exactly where the atoms from both slabs are connected. The nomenclature adopted in the present study is as follows, considering if atoms A and B belong to $Co_2MnAl$, and atoms C and D belong to MgO or GaAs:- for $Co_2MnAl$/MgO heterostructures, the "AB-CD" interfacial structure indicates that atom-A bonds with atom-C and atom-B with atom-D in a *top-to-top* arrangement. In contrast, for $Co_2MnAl$/GaAs heterostructures, the "A-C" interfacial configuration represents that the atom-A is connected to atom-C in a *top-to-top* configuration, whereas "AB-C" interfacial configurations signify that the C atoms connect at the fourfold hollow sites or Bridge sites between atoms-A and -B, as shown in Figure 1. For making the bridge configurations clearly distinguishable in the paper, they are intended to be indicated with the "(B)" suffix, for instance, "CoCo–Ga(B)" implies that the Ga atom from the GaAs layer is connected at the bridge site formed between two Co atoms of the $Co_2MnAl$ slab.

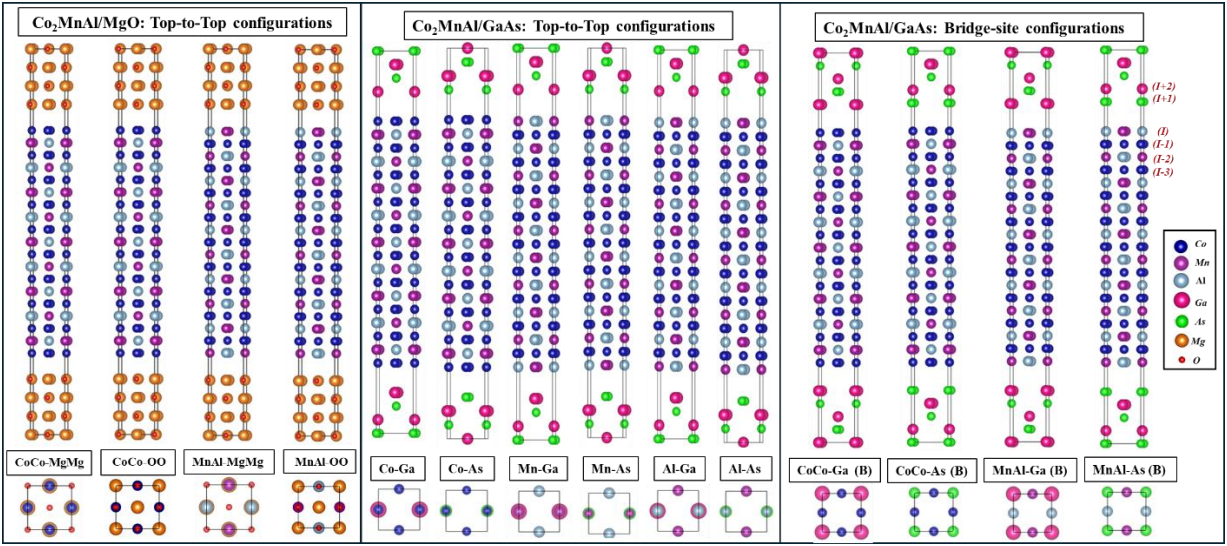

**Figure 1:** Various interfacial geometries for the $Co_2MnAl$/MgO and $Co_2MnAl$/GaAs heterostructures. The top row displays the side views of the interfacial geometry, while the bottom row illustrates the corresponding top views, when viewed along the vertical (out-of-plane) direction. The labels associated with each image (in the middle row) represent the nomenclature used for that particular interfacial morphology, which is used to identify different interfacial configurations. In particular, the letter *B* in parentheses denotes the interfacial **bridge-site configurations** for the $Co_2MnAl$/GaAs heterostructure. The adopted layer indices for the heterostructures and different atomic symbols are indicated in the figure's right corner.

In the present study, four *top-to-top* interfacial morphologies for (100)-$Co_2MnAl$/MgO heterostructure are considered. Whereas within (100)-$Co_2MnAl$/GaAs heterostructure, six *top-to-top* and four Bridge site interfacial morphologies are examined. All these morphologies are listed in Table 1 and shown in Figure 1. Here, the different heterostructure layers are labeled as shown in Figure 1. Since different studies might adopt different nomenclatures for defining interfaces, thereby, a precaution must be exercised when referring to the interfacial structures. At present, we cannot exclude the existence of other possible interface configurations with the given structural orientations of the constituting layers; however, we are limited by the computational resources. *Additionally, it should be noted very carefully that any reference to the "bulk" state of Co, Mn, Al, Mg, O, Ga, or As throughout this work refers to their state within the respective bulk compounds - $Co_2MnAl$, MgO, and GaAs - and not to the elemental bulk forms such as bulk-Co, bulk-Mn, bulk-Al, bulk-Mg, bulk-O, bulk-Ga, and bulk-As.*

*For modelling the heterostructures,* 19 monolayers (MLs) for $Co_2MnAl$, 7 MLs for GaAs/MgO for all *top-to-top* configurations are considered, while 9 MLs for GaAs are used for simulating bridge site configurations in (100)-Co$_2$MnAl/GaAs heterostructures. Notably for the Bridge site configurations, a slightly larger number of layers (9ML) is chosen to retain the identical atomic terminations at both interfaces, due to the symmetry of Co$_2$MnAl and GaAs slabs. The resulting heterostructures have a thickness of ~2–3 nm for the Co$_2$MnAl slab and ~1 nm for the X (MgO or GaAs) slabs, aligning with the typical technological application regimes of HA/NM heterostructures. The chosen thickness is also comparable to the previous computational studies [26–28] and ensures that the central layers' atoms in the heterostructures exhibit *nearly* bulk-like properties in terms of atomic magnetic moments (AMMs) and spin polarization. *Also, on further increasing the slab thicknesses, the change in the interfacial bond lengths and near-interface AMMs is impalpable* (*not shown for brevity*). All these rationales demonstrate that appropriate and experimentally relevant thicknesses for the composing layers are selected for studying the physical properties of the interfacial geometries.

For lattice matching of the different slabs in the heterostructures supercells, the chosen in-plane lattice parameters are ($a_{Co2MnAl}/\sqrt{2}$, $a_{Co2MnAl}/\sqrt{2}$) and ($a_{MgO}$, $a_{MgO}$) for Co$_2$MnAl and MgO slabs, respectively, for modelling the (100)-Co$_2$MnAl/MgO heterostructures. Similarly, for constructing the Co$_2$MnAl/GaAs-(100) heterostructures, the chosen in-plane lattice parameters for Co2MnAl and GaAs slabs are ($a_{Co2MnAl}$, $a_{Co2MnAl}$) and ($a_{GaAs}$, $a_{GaAs}$), respectively. The final in-plane lattice parameters of the heterostructures are calculated as the *average of the in-plane lattice parameters* of the constituent slabs (Co$_2$MnAl and X). The resulting heterostructures have a small lattice misfit of ≤ 6%, which minimizes the interface's defect density and facilitates the formation of the nearly epitaxial interfacial geometries. Furthermore, to enhance the computational efficiency, the heterostructures are reduced to a (1×1) in-plane supercell.

**Interfacial structural properties:**

Before calculating the physical properties of interfacial geometries, the structural optimization for the heterostructure is performed by relaxing both the lattice lengths and atomic positions along the (001) direction. Specifically, atoms within the five MLs of the Co$_2$MnAl HA and the two MLs of the adjacent material X (MgO or GaAs), closest to the interfaces, are allowed to relax. This relaxation minimizes interatomic forces and residual stress arising from interface formation. The structural optimization around the interfacial region is crucial, as it often leads to significant atomic reconstructions – such as uneven atomic layers, notable deviations in the average interplanar distances (either elongation or contraction), and variation in the interfacial geometries' specific bond length – relative to their bulk ideal configurations. Interestingly, the relaxation results in the present study reveal that, for any given atomic termination of Co$_2$MnAl slab in the optimized heterostructure, the interfacial geometries with MgMg (Ga) termination for the MgO (GaAs) slabs have a greater bond length than those with the OO (As) termination for the MgO (GaAs) slabs. This trend is somewhat similar to that observed for Ti$_2$MnAl in the Ti$_2$MnAl/MgO heterostructure [29]. This variation in the bond length across the interfacial geometries arises from the differing interatomic forces between the HA' atoms (Co, Mn, Al) and adjacent layers' atoms (Mg, Ga, O, As), which result in stronger repulsion interactions between the HA-atoms and Mg (Ga) atoms, compared to those between the HA' atoms and O(As) atoms. Microscopically, these interfacial-geometry-dependent interatomic forces stem from the modified ionic potential-energy landscape across the interface (which arises due to the different local bonding environment), and from the asymmetric atomic charge redistribution among the near interfacial atoms (which is driven by their different electronegativities). These combined effects lead to distinct interatomic forces in the vicinity of the interfacial region depending on the interfacial geometries, and ultimately lead to not only variation in bond lengths, but also to geometry-dependent adhesion energies, and atomic reconstruction (which often manifests as uneven atomic layers near the interface). Notably, for many configurations, the interfacial bond lengths are comparable to the interplanar distances of $Co_2MnAl$ (1.14 Å), MgO (1.41 Å), or GaAs (1.15 Å), *which indicates that these relaxed structures are physically reasonable and reliable.* Concerning the unevenness of atomic layers within heterostructures, atomic reconstructions are not very effective, leaving the near-interface layers nearly even.

**Table 1:** The adhesion energy and equilibrium bond for the optimized interfacial structures. If A, B atoms belong to Co2MnAl and C, D atoms belongs to MgO (or GaAs), then the "(100) AB-CD" interfacial structure means that at the interface, the atom A is connected to atom C, while the atom B is connected to atom D, whereas for the "A-C" interfacial configuration, the atom A is connected to atom C, all in the *top-to-top* configurations. On the other hand, for the bridge site interfacial configurations within (100)-$Co_2MnAl$/GaAs heterostructure, "AB-C" interfacial configurations signify that the C atoms connect at the fourfold hollow sites or Bridge sites between atoms A and B.

| Interfacial Structure | Bond | Bond length $d_{int}$ (Å) | Adhesion energy $\gamma$ (meV/A²-atom) |
|---|---|---|---|
| **(100) CoCo-MgMg** | Co-Mg | 3.27 | 0.77 |
| **(100) CoCo-OO** | Co-O | 2.29 | 1.3 |
| **(100) MnAl-MgMg** | Mn-Mg (Al-Mg) | 3.38 (3.49) | 0.55 |
| **(100) MnAl-OO** | Mn-O (Al-O) | 2.92 (3.10) | 0.37 |
| **(100) Co-Ga** | Co-Ga | 2.47 | 3.94 |
| **(100) Co-As** | Co-As | 2.31 | 4.29 |
| **(100) Mn-Ga** | Mn-Ga | 2.76 | 2.92 |
| **(100) Mn-As** | Mn-As | 2.66 | 3.03 |
| **(100) Al-Ga** | Al-Ga | 2.76 | 3.9 |
| **(100) Al-As** | Al-As | 2.68 | 3.09 |
| **(100) CoCo-Ga (B)** | Co-Ga | 1.77 | 4.02 |
| **(100) CoCo-As (B)** | Co-As | 1.23 | 7.05 |

| (100) MnAl-Ga (B) | Mn-Ga | 2.62 | 3.08 |
| (100) MnAl-As (B) | Mn-As | 1.55 | 3.32 |

Following this structural optimization, to compare the relative stabilities of the interfaces, the adhesion energies ($\gamma$), are calculated using the following equation:

$$\gamma = \left(E_{Co_2MnAl} + E_X - E_{Co_2MnAl/X}\right)/2A \tag{1}$$

Here, $E_{Co_2MnAl/X}$ and $A$ denote the total energy and total interfacial area for $Co_2MnAl/X$ heterostructures, respectively. Furthermore, $E_{Co_2MnAl}$ and $E_X$ represent the total energies for isolated $Co_2MnAl$ and X slabs within the same heterostructure supercell, when one slab is retained, and the other is replaced by the vacuum. Further, *the factor of 2 in the denominator accounts for the two identical interfaces in each heterostructure*. According to equation 1, the adhesion energy quantifies the ideal work required to separate the interface into two different surface slabs ($Co_2MnAl$ and X), thus serving as a measure of interface stability. A more positive adhesion energy means that the interface model is more energetically favorable or stable. The calculated interfacial adhesion energies are presented in Table 1. Among them, the (100) CoCo–OO geometry in the $Co_2MnAl/MgO$ heterostructure and the (100) CoCo–As (B) geometry in the $Co_2MnAl/GaAs$ heterostructure exhibit the highest adhesion energies, suggesting they are the most stable. The relative stability of the other configurations decreases in the following order: (100) CoCo–OO > (100) CoCo–MgMg > (100) MnAl–MgMg > (100) MnAl–OO, corresponding to decreasing adhesion energies as summarized in Table 1. Similarly, among the $Co_2MnAl/GaAs$ heterostructures, the order of increasing adhesion energy (and hence increasing interfacial stability) is: (100) Mn-Ga < (100) Mn-As < (100) MnAl-Ga (B) < (100) Al-As < (100) MnAl-As (B) < (100) Al-Ga < (100) Co-Ga < (100) CoCo-Ga (B) < (100) Co-As < (100) CoCo-As (B). In conclusion, it can be summarized that the interfacial morphologies compromising the CoCo- or Co-terminations for the $Co_2MnAl$ slab exhibit the larger adhesion energy as compared to those with MnAl- or Mn- or Al-termination of the $Co_2MnAl$ slab. As indicated in Table 1, adhesion energy is inversely related to bond length, consistent with stronger bonding interactions resulting in shorter distances - resembling previously reported trends for HAs [30]. Furthermore, for the $Co_2MnAl/GaAs$ heterostructures, bridge-site interfacial geometries display higher adhesion energies compared to top-to-top (direct facing) configurations. This can be attributed to smaller orbital overlaps (compared to the *top-to-top* configurations) and consequently reduced interfacial repulsion between the interfacial atoms of the two slabs.

**Interfacial electronic and magnetic properties:**

We now discuss the electronic and magnetic properties of the various interfacial geometries considered in the present study. To begin, the electronic behavior of the heterostructures is analyzed by calculating the projected density of states (PDOS) for atoms from both constituent slabs, *i.e.*, $Co_2MnAl$ and MgO or GaAs. When comparing the PDOS of different atoms in the heterostructures to that of the corresponding bulk state (*i.e.*, PDOS in bulk $Co_2MnAl$, MgO, or GaAs), it is observed that for heterostructure layers far from the interface, the constituting atoms' PDOS is nearly identical to those in bulk $Co_2MnAl$ and MgO or GaAs. This indicates that the interface formation has a negligible influence in the far-away region from the interface. In contrast, the atoms located near the interfacial region display significant deviations in their PDOS compared to those in the respective bulk state (bulk $Co_2MnAl$, GaAs, or MgO), clearly reflecting notable changes in the local electronic environment induced by the interface.

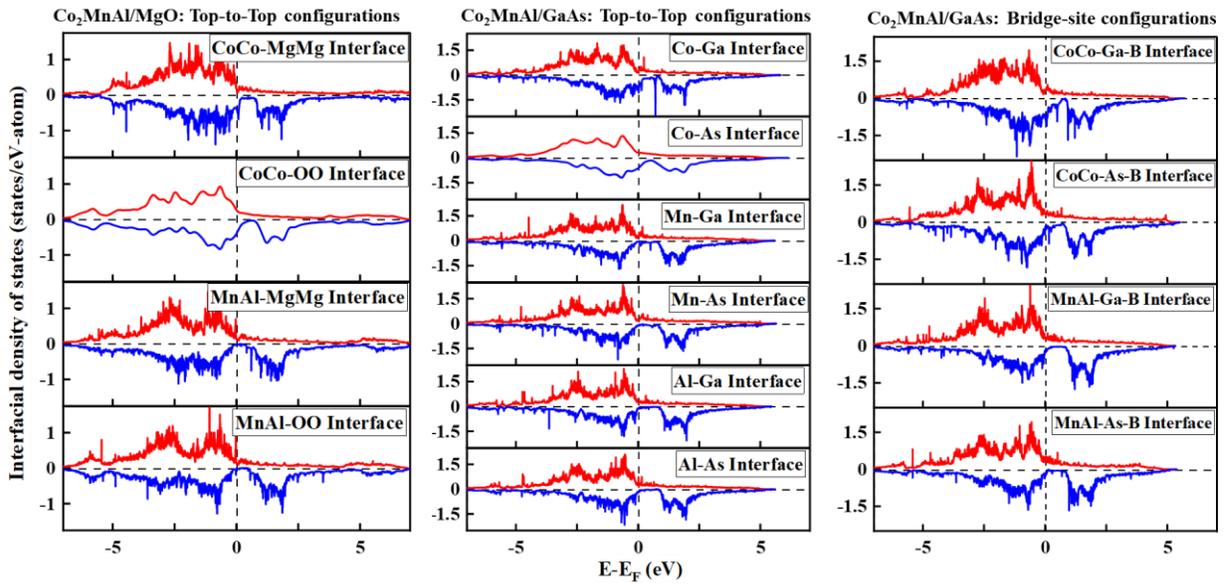

**Figure 2:** Spin-resolved interfacial density of states (states/eV-atom) for various $Co_2MnAl$/MgO and $Co_2MnAl$/GaAs heterostructures, projected across seven layers of the heterostructure - comprising five atomic layers from the $Co_2MnAl$ slab and two atomic layers from the X (MgO or GaAs) slab, as the interfacial effects are significant only within these seven layers. The Fermi level is set to zero energy, and the red and blue lines correspond to spin-up and spin-down projected DOS, respectively.

Notably, these changes are most pronounced in the interfacial atomic layers and gradually diminish with increasing distance from the interface. The corresponding PDOS for atoms within the heterostructures, along with comparisons to the respective bulk materials, are presented in Figures S1-S3 of the Supplemental Material. In Figures S1-S5, the atomic layer indexing to describe the heterostructure geometry is the same as illustrated in Figure 1. All these PDOS variations stem from the atomic relaxations of the near-interfacial atoms, which, in turn, alter the electronic exchange interactions for these atoms and, consequently, result in the modified PDOS, as shown in Figures S1-S4. Owing to these changes in PDOS of the near-interface layers' atoms, the electronic nature of the heterostructure evidently differs from the electronic nature of the constituting slab. To summarize these electronic nature changes, and to more clearly identify the overall electronic nature of the interfacial geometries, the spin-resolved DOS projected around the interfacial region is plotted in Figure 2, considering the five atomic layers for $Co_2MnAl$ and two atomic layers of MgO or GaAs (*i.e.*, by summation of PDOS atoms located between the (*I+2*) to (*I-2*) heterostructure layer). As these atomic layers capture all the interface formation influences on PDOS of heterostructure atoms, the DOS plot in Figure 2 can also be referred to as the interfacial DOS plot for the corresponding interfacial geometries. Here, it is noticeable that the resulting DOS for interfaces closely resembles the DOS of bulk-$Co_2MnAl$, with some minor changes. This similarity arises because these interfacial DOS primarily derive from the five $Co_2MnAl$ layers used in plotting the interfacial DOS, while only two atomic layers are present in MgO or GaAs. Furthermore, near the Fermi level ($E_F$), the DOS of $Co_2MnAl$ is significantly higher than that of MgO or GaAs, further reinforcing this resemblance. Also, it is worth noting that the interfacial DOS noticeably differs in the minority bands, whereas the majority band remains very similar to the bulk-$Co_2MnAl$ DOS for all structures. These changes are attributed to the efforts to minimize the band energies for the interfacial geometries, as even small changes in the majority DOS would involve larger energy changes for the corresponding interfacial geometries (majority spin DOS is considerably larger than those of minority spin DOS, as seen in Figure 2).

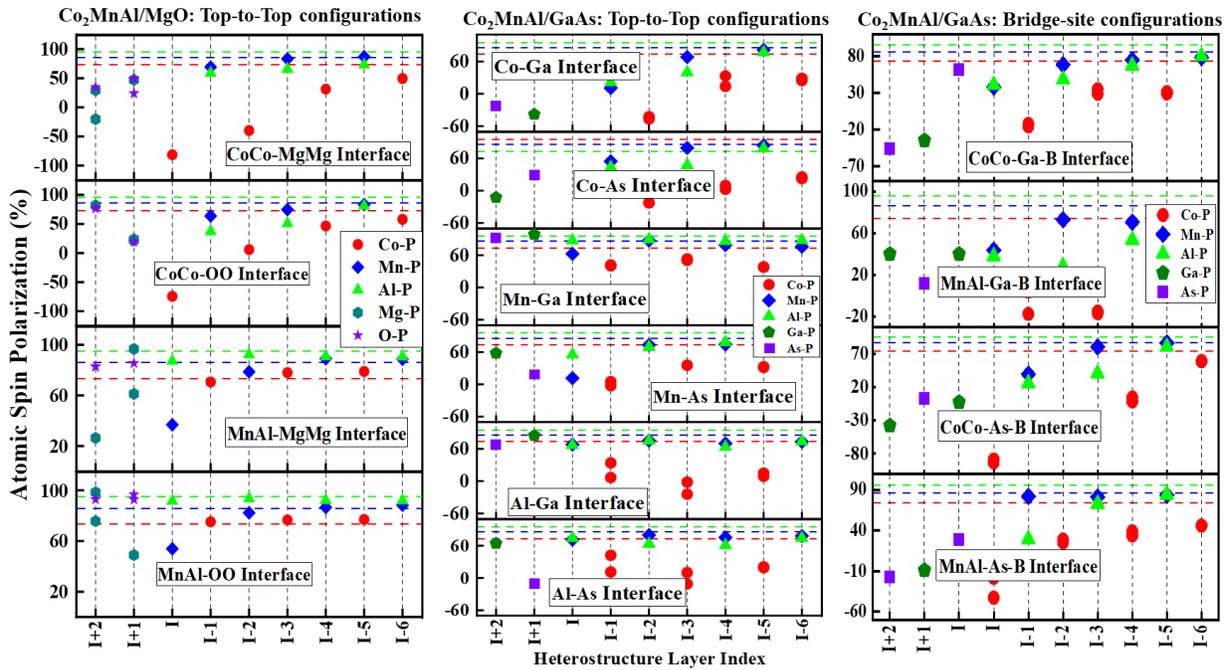

**Figure 3:**:Atomic-spin polarization (AP) for different $Co_2MnAl$/MgO and $Co_2MnAl$/GaAs heterostructures with respect to the heterostructure layer index. The red circles, blue diamonds, and green triangles represent the AP for Co, Mn, and Al atoms in the heterostructures, respectively. The corresponding APs in bulk $Co_2MnAl$ for Co, Mn, and Al are indicated by red, blue, and green dashed lines, respectively. In addition, the AP for Mg, O, Ga, and As atoms in the heterostructures is represented by basil hexagons, violet stars, basil pentagons, and violet squares, respectively. The corresponding bulk values for Mg, O, Ga, and As atoms are not shown here due to very small PDOS at the Fermi level.

Among all interfacial geometries, the interfaces composed of Co–Co terminations for the $Co_2MnAl$ slabs, a rich metallic character is observed due to the significant density of states (DOS) within the pseudogap. This strong metallic behavior appears to originate from the interfacial Co-atoms, with a minor contribution from the (*I-1*) layer, as evidenced by PDOS plots in Figures S1-S4. Whereas for the heterostructures, comprising the MnAl-terminations of the $Co_2MnAl$ slab, a nearly half-metallic nature is observed, as the interfacial atoms have similar PDOS as in bulk $Co_2MnAl$, MgO, or GaAs. Thereby, it can be concluded that the interfacial morphologies involving the MnAl-terminated (or Mn- or Al-terminated) $Co_2MnAl$ slab, such as - MnAl-MgMg, MnAl-OO, Mn-Ga, Mn-As, Al-Ga, Al-As, MnAl-Ga (B), and MnAl-As (B) - will possess a nearly half-metallic nature. Whereas, for the interfaces with CoCo-terminated (or Co-terminated) slab for $Co_2MnAl$, such as CoCo-MgMg, CoCo-OO, Co-Ga, Co-As, CoCo-Ga (B), and CoCo-As (B), a rich metallic behaviour will be observed. This clear distinction in interfacial electronic character implies that these two classes of interfacial geometries will exhibit markedly different responses in experimental studies, particularly in spintronic device performance, where metallicity versus half-metallicity plays a critical role.

To quantify the interface formation effect on the electronic nature of heterostructures, both total interfacial spin polarization ($P_{inf}$) and atom-resolved spin polarization (AP) at the $E_F$ for all heterostructures are also quantified, using the standard DOS spin polarization formula $P = \frac{D_1-D_2}{D_1+D_2} \times 100$. Here, $D_1$ and $D_2$ denote the spin-up and spin-down DOS at $E_F$. For the interfacial spin polarization calculation, the interfacial DOS is utilized, while the atom-resolved spin polarization is extracted from the PDOS of individual atoms. The computed interfacial spin polarization values for the various interfacial geometries with Co2MnAl/MgO heterostructures are as follows: CoCo-MgMg: -32.88%, CoCo-OO: -20.00%, MnAl-MgMg: 77.97%, MnAl-OO: 79.59%. Similarly, for the $Co_2MnAl$/GaAs heterostructures, the interfacial spin polarization values for different interfacial geometries are: -60.86% (Co-Ga), -33.51% (Co-As), 66.15% (Mn-Ga), 32.96% (Mn-As), 30.11% (Al-Ga), 35.07% (Al-As), 39.81% (CoCo-Ga), 17.77% (CoCo-As), 17.98% (MnAl-Ga), 19.56% (MnAl-As). These results confirm that interfaces with MnAl-, Mn- or Al-terminations for $Co_2MnAl$ slab retain a high degree of spin polarization - consistent with their nearly half-metallic nature, while those with CoCo-or Co-terminated $Co_2MnAl$ slab exhibit significantly reduced or even negative spin polarization, indicative of metallic behavior. The atomic spin polarizations for different interfaces are illustrated in Figure 3. The calculated atomic spin polarization directly correlates with the PDOS at the $E_F$ for respective atoms. Consequently, the spin polarization of the heterostructures' atoms also changes based on their PDOS variations. Figure 3 presents the atomic spin polarization only from (*I*+2) to ($I$-5) layers, *i.e.*, including two layers of MgO or GaAs slab and five layers of Co$_2$MnAl slab; as if further going deeper into the heterostructures along any slab, bulk-like atomic spin polarization is observed. As seen in Figure 3, the atomic spin polarization of the near-interface atoms is consistently reduced across all heterostructures studied, with the extent depending on the interfacial geometry.

For the interfacial geometries comprising the MnAl-terminations (or Mn- or Al-terminations) for the Co$_2$MnAl slabs, there is a slight reduction in atomic polarization. Whereas for the interfacial geometries with the CoCo-terminations (or Co-terminations) for the Co$_2$MnAl slab, this reduction is significantly larger, even leading to the inverted spin polarization than those near the central layers of heterostructures (negative atomic spin polarization, due to larger spin down DOS compared to spin up DOS at the Fermi level). Another noteworthy observation is that for all interfacial geometries with *top-to-top configurations*, the atomic spin polarization for chemically identical atoms within one heterostructure layer is the same; thereby, among the chemically identical atoms, the atomic spin polarization for only one atom is shown in Figure 3. In contrast, for the bridge-site configurations, even for chemically identical interfacial atoms, the atomic spin polarization differs due to variations in the local bonding environment. This asymmetry arises because of the inequivalent bonding geometry in bridge-site configurations for identical atomic species. Thereby, the atomic spin polarization for both atoms is displayed in Figure 3. A similar approach is followed in the description of atomic magnetic moments (AMMs), which are presented in Figure 4 and discussed in the following paragraphs.

Another important observation is the spatial extent of spin polarization variation across heterostructure layers for different configurations. The spread of atomic spin polarization modifications extends up to the ($I$–1), ($I$–2), and ($I$–3) layers along the Co$_2$MnAl slab direction for the Co$_2$MnAl/MgO heterostructures, Co$_2$MnAl/GaAs heterostructures (top-to-top configurations), and Co$_2$MnAl/GaAs heterostructures (Bridge site configurations). While it is observed that notable changes in the projected density of states (PDOS) are confined primarily to the (I–1) layer, and the variations in atomic spin polarization stem from these PDOS modifications, one might expect that changes in atomic spin polarization would also be limited to the (I–1) layer across all interfacial geometries. However, it is important to recognize that the calculated spin polarization is highly sensitive to the DOS due to its formulation. As a result, even subtle differences—often imperceptible in the PDOS comparison—can induce noticeable variations in atomic spin polarization. Therefore, variations in spin polarization are observed across multiple atomic layers. Nevertheless, it should be emphasized that such minor changes in atomic spin polarization do not necessarily influence the overall interpretation of the electronic nature of the heterostructures.

Next, let us discuss the magnetization of interfacial geometries. The AMMs for heterostructure layers can also be understood quantitatively from the atomic DOS, as the spin magnetic moment is proportional to the difference between the total spin-up and spin-down electrons. Given that the spin-resolved PDOS for the near-interface atoms differs from the bulk state manifested by the impact of interface formations, thereby, the change in atomic magnetic AMMs is also anticipated for the near-interface atoms. These changes in the AMMs stem from the same origin that gives rise to PDOS variations and are discussed in the last paragraph. As shown in Figure 4, notable changes in AMMs are observed for all interfacial geometries, and all heterostructures show enhanced AMMs compared to those in the bulk state. Also, the enhancement level is similar for all heterostructures. For the atoms away from the interface, from (I-2) layers of the heterostructures, bulk-like AMMs are observed.

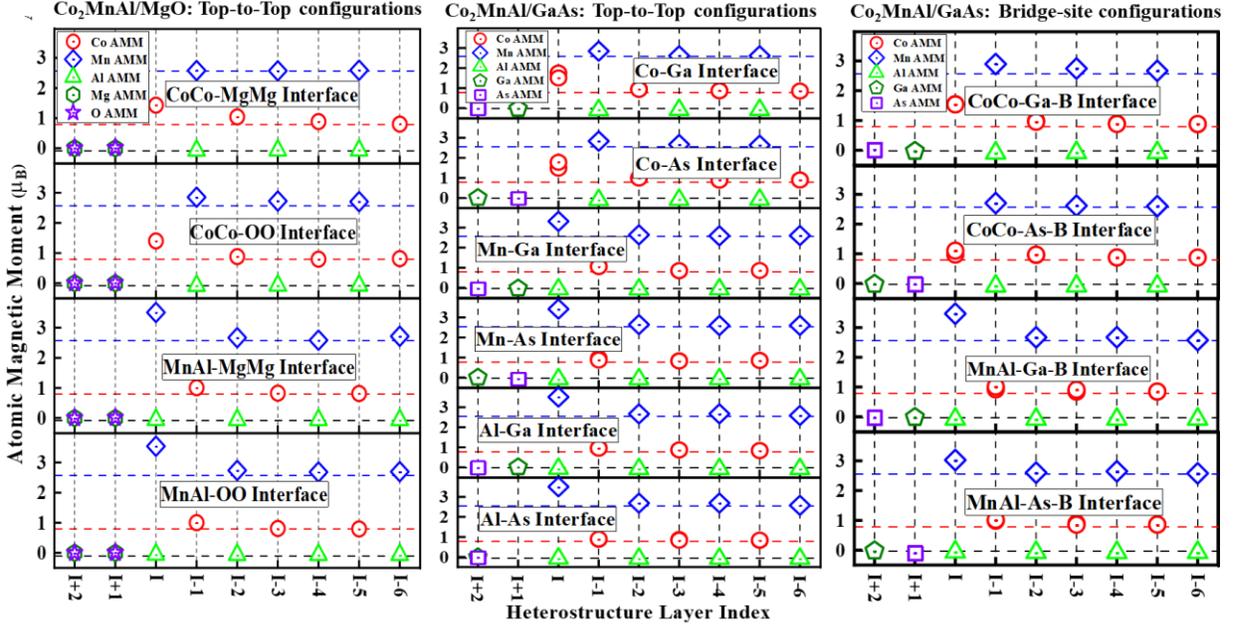

**Figure 4**: Atomic-resolved atomic magnetic moments (AMMs) with respect to the heterostructure layer index. The red circles, blue diamonds, and green triangles represent the AMMs for Co, Mn, and Al atoms in the heterostructures, respectively. The corresponding AMMs in bulk $Co_2MnAl$ for Co, Mn, and Al are indicated by red, blue, and black dashed lines, respectively. In addition, the AMMs for Mg, O, Ga, and As atoms in the heterostructures are represented by basil hexagons, violet stars, basil pentagons, and violet squares, respectively. The corresponding bulk values for these atoms (in bulk MgO and GaAs) are shown as black solid lines, similar to the bulk Al AMM in $Co_2MnAl$, due to their similarly negligible magnetic moments.

## IV. Conclusion

The structural, electronic, and magnetic properties of various interfacial geometries stemming from (100)-$Co_2MnAl$/MgO and (100)-$Co_2MnAl$/GaAs heterostructures have been studied using the plane wave pseudopotential-based density functional theory calculations. For $Co_2MnAl$/MgO heterostructures, the CoCo-MgMg, CoCo-OO, MnAl-MgMg, and MnAl-OO interfacial geometries, all with the *top-to-top* (direct facing) configurations for the interfacial atoms, are considered. On the other hand, for the Co$_2$MnAl/GaAs heterostructures, the following interfacial geometries: Co-Ga, Co-As, Mn-Ga, Mn-As, Al-Ga, and Al-As (with the *top-to-top* configurations), and CoCo-Ga (B), CoCo-As(B), MnAl-Ga (B), MnAl-As (B) (with *bridge site configurations*) are considered. The Interfacial geometries with CoCo-terminated ends for the Co$_2$MnAl slab—such as *CoCo-MgMg, CoCo-OO, Co-Ga, Co-As, CoCo-Ga(B), and CoCo-As(B)*—exhibit larger adhesion energies compared to those with Mn-, Al-, or MnAl-terminated ends for the Co$_2$MnAl slab, indicating a higher stability for them, in both Co$_2$MnAl/MgO and Co$_2$MnAl/GaAs heterostructures. Specifically, for the Co$_2$MnAl/GaAs heterostructures, bridge-site interfacial geometries display higher adhesion energies compared to the *top-to-top* (direct facing) configurations. This is attributed to smaller orbital overlaps (compared to the *top-to-top* configurations) and consequently reduced interfacial repulsion between the interfacial atoms of the two slabs. Concerning the electronic nature of the interfacial geometries, the interfacial geometries with MnAl- or Mn-, or Al-termination for the Co2MnAl slab, exhibit an electronic structure close to a half-metallic nature, and resemble that of the bulk Co$_2$MnAl electronic nature. In contrast, the interfacial geometries with CoCo- or Co-termination for the Co$_2$MnAl slab exhibit a strongly metallic character. In relation to magnetization, all studied interfacial morphologies exhibit enhanced magnetic moments compared to those in the bulk geometries in Co$_2$MnAl, GaAs, or MgO. These variations in the electronic nature and magnetization stem mainly from interfacial atoms and their first nearest neighbors.

## Acknowledgements

PARAM Rudra, a national supercomputing facility at Inter-University Accelerator Centre (IUAC), New Delhi, has been used to obtain the results presented in this paper. A.K. acknowledges the Council of Scientific and Industrial Research (Grant No. 09/086(1356)/2019-EMR-I), India, for the senior research fellowship.

# Supplemental Material

**Interfacial atomic density of states (Interfacial PDOS):** The changes in the electronic and magnetic properties for the interfacial geometries primarily originate from the interfacial atoms.

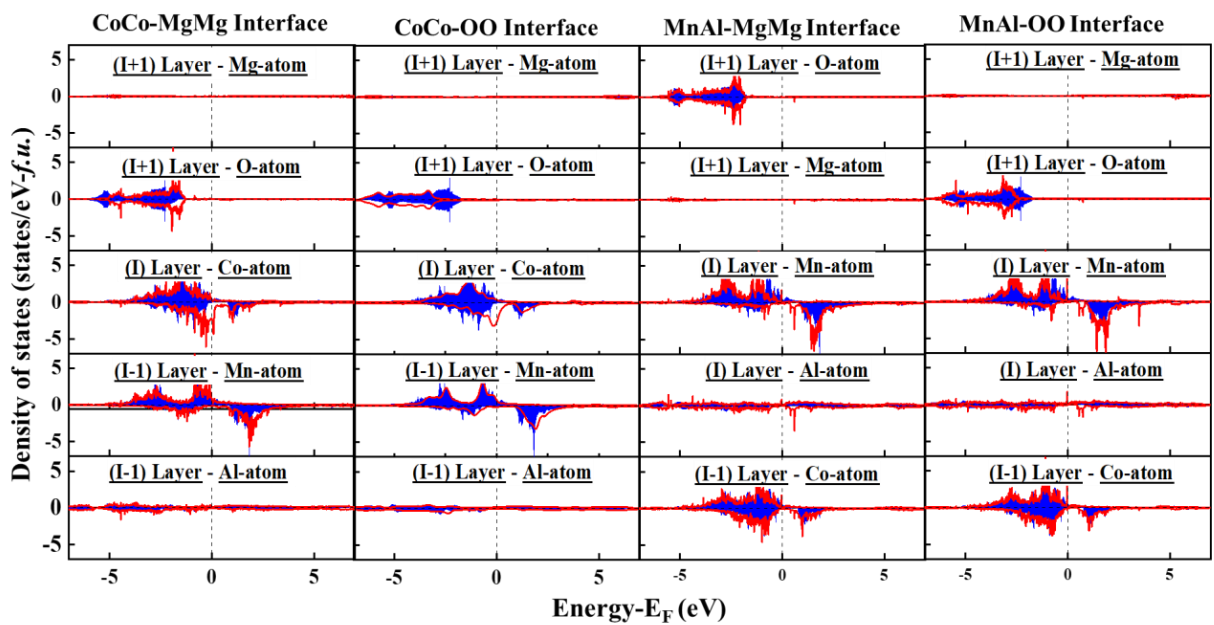

**Figure S1**: Partial density of states (PDOS) for the near-interfacial atoms for different interfacial geometries within $Co_2MnAl/MgO$ heterostructures, with Top-to-Top configurations.

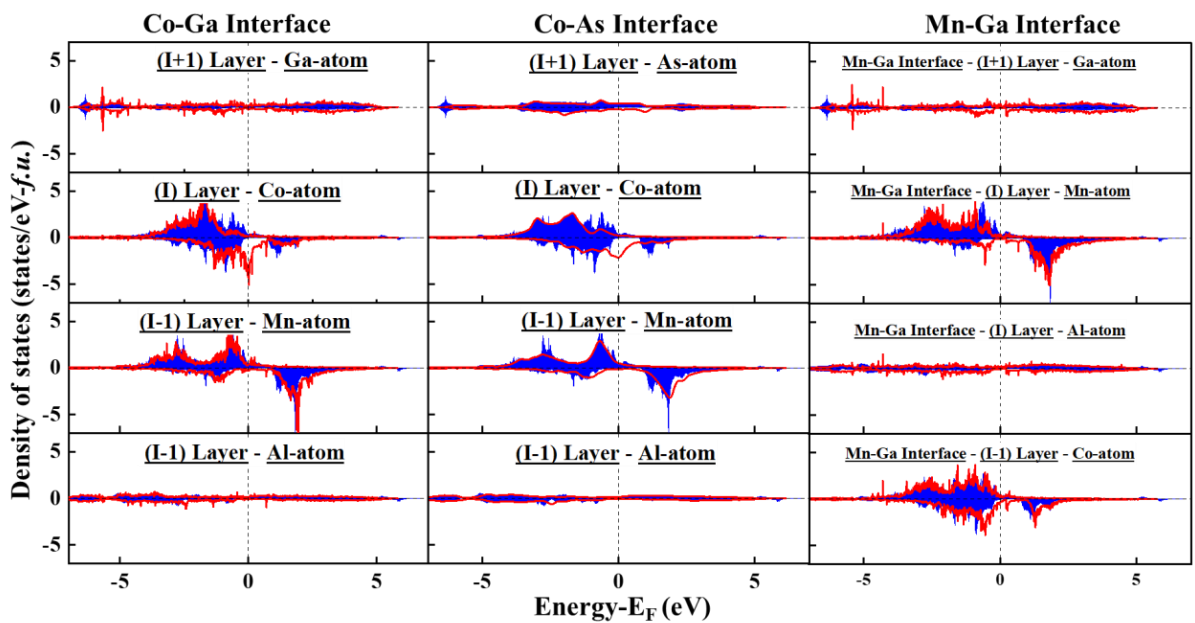

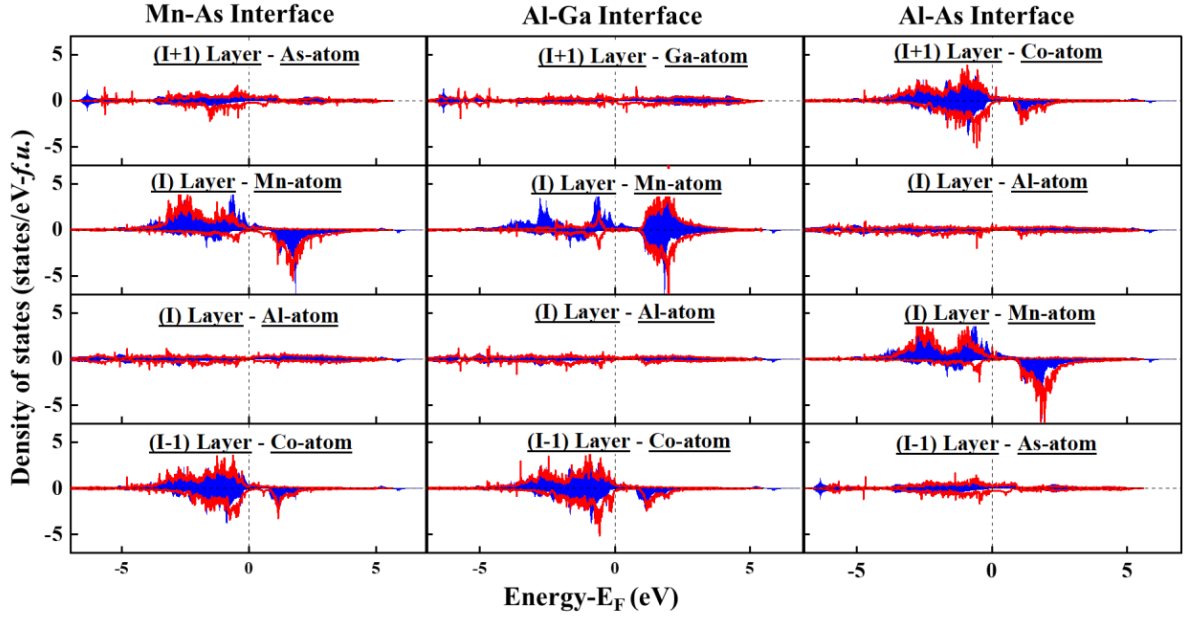

**Figure S2**: Partial density of states (PDOS) for the near-interfacial atoms for different interfacial geometries within $Co_2MnAl/GaAs$ heterostructures, with Top-to-Top configurations.

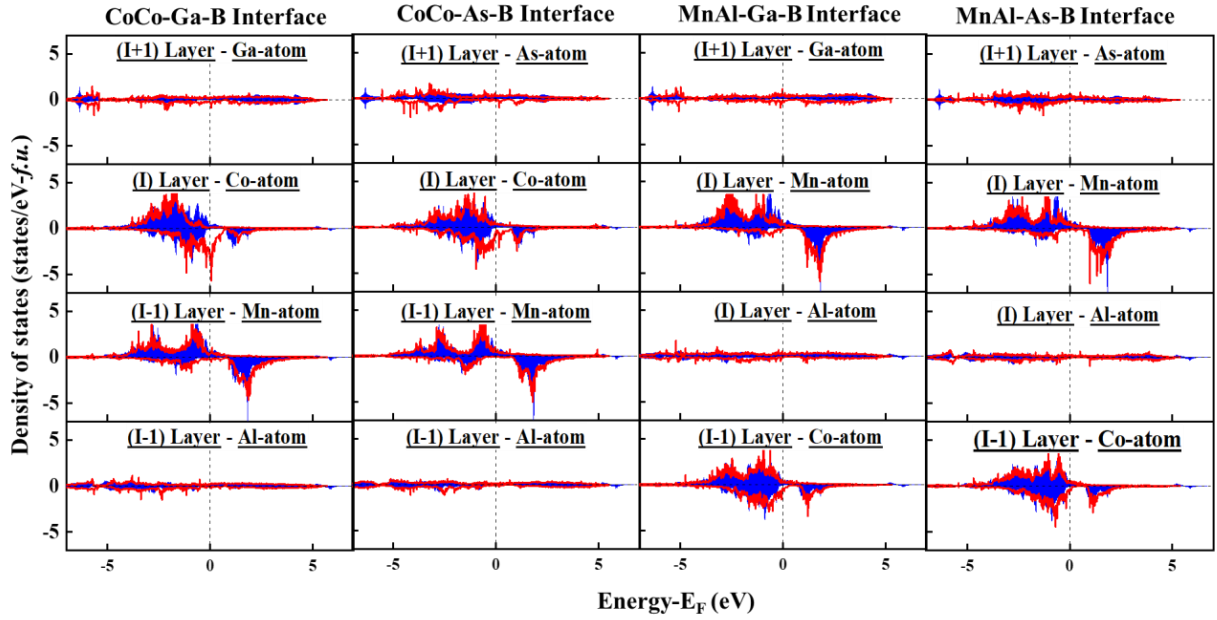

**Figure S3**: Partial density of states (PDOS) for the near-interfacial atoms for different interfacial geometries within $Co_2MnAl/GaAs$ heterostructures, with Bridge site configurations.